## 3.4. Atomic vibrations in glasses

**Bernard Hehlen and Benoît Rufflé**

Laboratoire Charles Coulomb (L2C), UMR 5221 CNRS, Université de Montpellier, 34095 Montpellier, France

**Abstract**

In glasses, atomic disorder combined with atomic connectivity makes understanding of the nature of the vibrations much more complex than in crystals or molecules. With a simple model, however, it is possible to show how disorder generates quasi-local modes on optic branches as well as on acoustic branches at low-frequency. The latter modes, possibly hybridizing with low-lying optic modes in real glasses, lead to the excess, low-frequency excitations known as "boson-peak modes", which are lacking in crystals. The spatially quasi-localized vibrations also explain anomalies in thermal conductivity and the end of the acoustic branches, two other specific features of glasses. Together with the quasi-localization of the modes at the nanometric scale, structural disorder lifts the crystalline or molecular spectroscopic selection rules and makes interpretation of experiments difficult. Nevertheless, vibrations in simple glasses such as vitreous silica or vitreous boron oxide are nowadays rather well described. But a comprehensive understanding of the boson peak modes remains a highly debated issue as illustrated by three archetypal glass systems, vitreous $SiO_2$ and $B_2O_3$ and amorphous silicon,

# 1. Introduction

Atoms have by definition essentially fixed position in a solid. Their only degrees of freedom are thus vibrations around these positions with amplitudes that increase with temperature. As a thermodynamic measure of these changes, the heat capacity then determines the temperature dependence of the internal energy, entropy and other functions. But the relevance of atomic vibrations is not restricted to thermal properties because thermal energy is involved in any process where potential energy barriers must be overcome. This is the case of not only any kind of phase transformations, including magnetic or ferroelectric transitions, but also of transport of heat, electrically charged species or atoms.

In a crystal, atoms do not vibrate independently from one another. Vibrations are instead described as a set of harmonic plane waves termed *phonons* whose energies are quantized, whose number and type are determined by the symmetry of the lattice, and whose frequencies vary with the wave vectors according to specific dispersion relations. Although the fundamental basis of *lattice dynamics* was laid down by Born and von Karman as early as 1912, the lack of spatial symmetry prevents this theory from being applied to amorphous solids. The problem of atomic vibrations in glasses has thus long been laid aside until the impressive development of new experimental techniques, computing capabilities and theoretical advances has made it possible to tackle it fruitfully in the last past decades.

The connection between the macroscopic and microscopic aspects of atomic vibrations is most readily established through the *vibrational density of states*, $g(\Omega)$, which is the distribution of the number of vibrational modes as a function of the frequency $\Omega$. The vibrational isochoric heat capacity $C_v(T)$ is directly derived from $g(\Omega)$ through

$$C_v(T) = \int_0^{\Omega_m} c_v(\Omega, T)\, g(\Omega)\, d\Omega, \qquad (1)$$

where $\Omega_m$ is the highest vibrational frequency. As for, $c_v(\Omega,T)$ it is the heat capacity of a single oscillator given by the Einstein function

$$c_v(x) = x^2 e^x/(e^x - 1)^2, \quad (2)$$

where $x = \hbar\, \Omega/k_B T$, and $k_B$ and $\hbar$ are Boltzmann and reduced Planck constants, respectively.

In the last past decades, much progress has been made either theoretically or experimentally to determine $g\,(\Omega)$ as well as the various vibrational excitations existing in glasses. These advances will thus be reviewed in this chapter with a focus on low-temperature conditions under which a rich phenomenology manifests itself. When heat capacities are approaching their Dulong-and-Petit limits, which is already the case for oxide glasses near room temperature, significant effects of structure and composition on vibrational properties in contrast necessarily tend to vanish. This is why such high-temperature conditions will not be considered in this chapter.

The starting point will be the classical description of phonons in perfectly ordered structures with a simple one-dimensional model. The impact of disorder on the vibrational properties of the system will reveal the loss of the plane-wave character of the normal modes and the concomitant apparition of quasi-localized modes. We follow with the comparison between the low-temperature thermal properties of crystalline and disordered solids, underlining the relation between the well-known anomalous behavior of the latter materials and their vibrational properties. Finally we discuss the implications for vibrational spectroscopy of sound waves and optic modes, including the boson peak.

## 2. Atomic vibrations in disordered solids

*2.1. The diatomic linear chain*

A simple system is a chain of $N$ atoms of mass $m$ alternating with $N$ atoms of mass $M$ to which they are connected by springs of force constant $K$. If periodic conditions are applied, the chain forms an infinite periodic lattice of unit cells comprising one of each mass. The solution to the equation for longitudinal motions is a set of well-defined propagating plane waves with eigenfrequencies $\Omega_j$ and associated wave vectors $Q_j$, which define two dispersion curves $\Omega_j(Q_j)$, namely, an acoustic and an optic branch representing in- and anti-phase motion of the masses within a unit cell, respectively. Each solution **j** is called a *normal mode* or an *eigenmode*. For the $j^{th}$ normal mode, the displacement of mass l of one species reads

$$U_{j,l}(t) = u_j \exp(i[Q_j r_l - \Omega_j t]), \qquad (3)$$

where $u_j$ is the amplitude of the mode and $r_l$ the position of the mass in the chain. At every moment, there exists a perfect spatial oscillatory pattern of the mass displacements $u_{j,l} = u_j \exp(iQ_j r_l)$, characterized by $Q_j$.

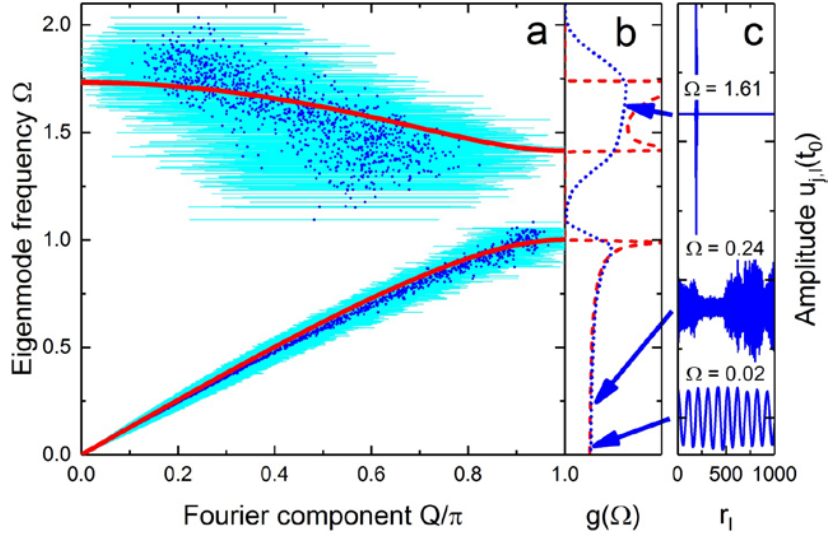

Fig. 1. Vibrational properties of a disordered diatomic linear chain. (a) Eigenmode frequencies $\Omega_j$ as a function of the mean value $\overline{Q}_j$ of the wave vector spectral density $\alpha_j(Q)$ (dots); standard deviations of $\alpha_j(Q)$ centered on $\overline{Q}_j$ characterizing the Q-spread of the $j^{th}$ eigenmode (very thin horizontal lines); dispersion curves of the acoustic and optic modes of the ordered chain (thick oblique curves). (b) Vibrational density of states $g(\Omega)$ of the crystalline (dashed line, $g_c(\Omega)$) and the disordered chains (dotted line, $g_d(\Omega)$). (c) Displacement snapshots for the three eigenmodes indicated.

One can then introduce disorder by varying the spring constants $K_i$ along the chain, the masses, or their equilibrium positions. Here we will assume a normal distribution of spring constants with a mean value $K_0$ and a standard deviation $\delta K$ to diagonalize the dynamical matrix and derive the eigenfrequencies and eigenmodes of the system. Owing to disorder, $U_{j,l}$ can no longer be simply expressed as in eqn (3) so that it is no longer possible to define $Q_j$ strictly. One can nevertheless expand $u_{j,l}$ in a Fourier series of the components $Q_k$ with amplitudes $\alpha_{j,k}$ [1]. For a quasi-plane wave normal mode **j**, the amplitude is significant only for $\alpha_{j,k}$ for which $\mathbf{j} \approx \mathbf{k}$. Conversely a localized mode has a large range of non-zero $\alpha_{j,k}$. The nature of a normal mode j can thus be roughly characterized by the mean value $\overline{Q}_j$ and the standard deviation of its wave vector spectral density $|\alpha_{j,k}|^2$. These coefficients also control the response of the system to a plane-wave vibrational excitation [1]. As an example, eigenfrequencies $\Omega_j$ have been computed for a disordered chain comprising $N = 2000$ masses as a function of $\overline{Q}_j$ (Fig. 1) with $\delta K/K_0 = 0.25$ and a mass ratio $M/m = 2$.

As noted long ago [2], the vibrational density of states of the disordered chain $g_d(\Omega)$ is not much affected by disorder, especially for the acoustic modes whose density is actually similar to that of a crystalline counterpart $g_c(\Omega)$. The main differences occur near termed frequencies where $g_c(\Omega)$ exhibits sharp maxima, which are termed *Van Hove singularities* and are smeared out by disorder (Fig. 1b). Although this similarity is holding particularly true for optic modes, $g_d(\Omega)$ is higher than $g_c(\Omega)$ by only about 3 % even at low frequencies, *i.e.* in the acoustic regime.

In contrast, the nature of the vibrational modes is strongly modified by disorder (Figs. 1a and 1b). The progressive scattering of the initially plane waves is clearly illustrated by the wide *Q*-spread of the eigenmodes (Fig. 1a). As generally found, the higher the frequency, the larger is the deviation from a plane-wave excitation. In the long-wavelength limit, the very low-frequency acoustic modes hardly differ from those of their crystalline counterparts. Details of the force constant distribution is indeed of little concern to these long-wavelength acoustic excitations as fluctuations are averaged out.

The eigenvectors [mass displacements] of these modes exhibit almost perfect oscillations in space as illustrated in Fig. 1c by the lowest line corresponding to $\Omega_j = 0.02$. This is no longer true for $\Omega_j = 0.24$ for which the envelope of the mass displacements shows important spatial variations. This mode is still an extended [collective] mode as all the masses participate in the vibration, albeit with different amplitudes. The loss of the plane wave character increases dramatically for modes at higher frequency as shown for $\Omega_j = 1.61$ belonging to the optic branch in the crystalline chain. Most of the vibrational amplitude is localized on a couple of neighboring masses. All high-frequency modes show significant displacements of neighboring masses only, occurring at randomly distributed spatial positions. But these modes are not truly localized because their vibrational amplitudes are extremely small for many masses, but not exactly zero.

*1.2. Real amorphous solids*

Scattering of the vibrational modes thus increases with frequency even when elastic disorder is small enough that the vibrational density of states is similar to that of a periodic lattice. To what extent such a simple picture can be generalized has been much debated because real glasses combine positional, mass and elastic disorder in complex 3-d structures made up of coupled structural entities. Further, disorder-induced mixing of transverse and longitudinal polarized excitations should arise at high frequency [1].

In the long-wavelength limit, sound waves propagate in glasses as they do in crystals. At that scale, amorphous solids are isotropic continuous elastic media whose low-frequency sound-waves can still be reasonably described as quasi plane-wave acoustic excitations. Phonon-like transverse acoustic excitations showing linear dispersion have been for example measured in vitreous silica [$v$-$SiO_2$] up to ~ 440 GHz [3], which corresponds to a length-scale of ~10 nm. This result might thus suggest that the low-temperature thermal properties of glasses should mimic those of their crystalline counterparts, at least below 2-3 K, but this does not seem to be generally the case. From the expected non plane-wave character of the high-frequency eigenmodes it is further anticipated that selection rules should be relaxed in spectroscopic studies, complicating the analysis and limiting the amount of information that can be obtained from vibrational spectra.

## 3. Vibrations and thermal properties

*3.1. Heat capacity*

The heat capacity $C_v \approx C_p$ of a perfect crystal obeys Debye law $C_p(T) = C_D T$ at temperatures below $\sim 0.01\ \theta_D$, where $\theta_D$ is the Debye temperature and $C_D \propto \theta_D^{-3}$ can be calculated from the sound velocities and the atomic density. At higher temperatures, $C_p$ begins to probe the details of the atomic structure via the acoustic dispersion curves throughout the Brillouin zone and eventually the optic modes.

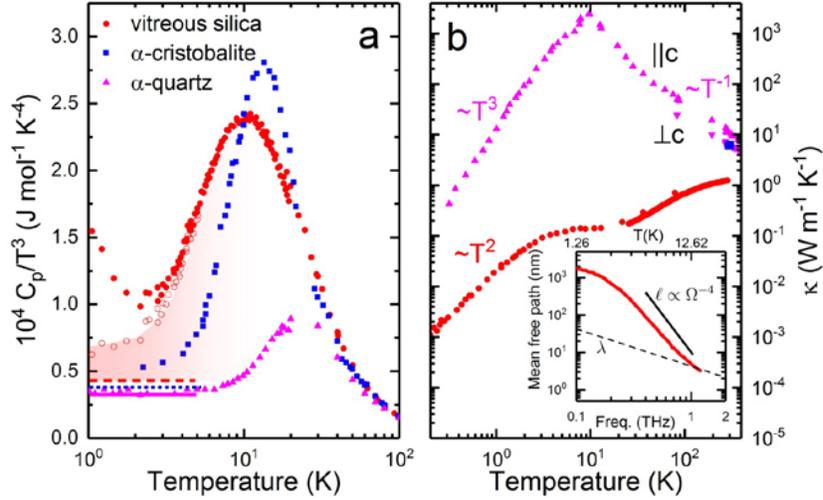

Fig. 2. Low-temperature thermal properties of $SiO_2$ phases. (a) Low-temperature heat capacity of vitreous silica, α-cristobalite and α-quartz in plots of $C_p/T^3$ against $T$. Horizontal lines: Debye constants (dashed, dotted and full line, respectively). Open circles: $C_p/T^3$ of vitreous silica without the Two-Level System contribution. (b) Thermal conductivity of the same polymorphs. Inset: frequency dependence of the mean free path in vitreous silica in the dominant-phonon approximation.

The measured heat capacity of α-quartz nicely illustrates this point. Below 5 K, $C_p/T^3$ almost perfectly matches the expected Debye value whereas the upturn above 10 K is fully described by the curvature of the acoustic phonon branches. In α-cristobalite, the $SiO_2$ polymorph whose density is close to that of the glass, $C_p/T^3$ reaches its constant Debye value only below 2 K, in agreement with its lower atomic density. Giving rise to a large bump around 13 K, the very steep increase at higher temperature is mostly related to the rapid flattening of the transverse acoustic branches in the <110> direction [4,5]. Because a given vibrational mode contributes to the heat capacity according to its relative weight in the vibrational density of states, the intense Raman active zone center mode of α-cristobalite around 50 cm$^{-1}$ [6] for example enhances $C_p$ from 15 K, just above the maximum of $C_p/T^3$.

In contrast, the heat capacity of $v$-SiO$_2$ does not conform to Debye model. Its larger value at 2 K indicates extra modes at low-frequencies. Close to 0 K, part of this excess is due to an almost linear contribution proportional to temperature, which is associated with tunneling states [7-9], whose discussion is beyond the scope of this chapter. Subtracting this feature (Fig. 2a) still leaves an excess over the Debye prediction. In agreement with their analogous local order and atomic packing, $v$-SiO$_2$ and cristobalite display similar peaks in $C_p(T)/T^3$ and reduced density of states $g(\Omega)/\Omega^2$. Such common features early indicated [4] that low-energy peaks are not a peculiarity of glasses [10,11] although they do not necessarily imply a common origin for both kinds of phases. The heat capacity of cristobalite is for instance completely understood in terms of phonon-branch dispersion. In contrast, the linear dispersion of sound waves up to at least 440 GHz [3] in $v$-SiO$_2$ gives a constant Debye value up to ~5.5 K (dashed line in Fig. 2a) which is inconsistent with both the position and width of the peak in $C_p(T)/T^3$. One must assume instead a $g(\Omega) \propto \Omega^4$ relationship on top of the Debye contribution [7]. Such an $\Omega^4$ dependence of the density of states is an ubiquitous feature in disordered systems [12]. But whether the excess above the Debye expectation at temperature below the maximum in $C_p/T^3$ (shaded area in Fig. 2a) is a manifestation of disorder and, as such, is related to a strong scattering of the acoustic phonons in glasses occurring at high frequencies is a central and still unsettled question. Only these vibrational excitations define the *boson peak modes*, as will be further discussed below.

*3.2. Thermal conductivity*

In dielectric crystals, significant scattering of acoustic phonons is reflected in the low-temperature thermal conductivity κ. With the standard kinetic equation for gases, one approximately describes the temperature dependence of κ by

$$\kappa = \frac{1}{3} C_v v l, \qquad (4)$$

where $C_v$ is the heat capacity per volume of the excitations providing the thermal transport, $v$ their velocity of propagation, and $l$ their mean free path [13]. This expression is obviously valid only for propagating phonons, *i.e.* below 1-2 THz in glasses as discussed below. From ambient, $\kappa$ increases with decreasing temperature through the increasing lifetimes of acoustic excitations. As there are fewer and fewer phonons, $l$ increases as $T^{-1}$ at high temperatures. The mean free path is of course limited by sample dimensions so that $\kappa$ goes through a maximum, the *Casimir limit*, and then decreases with the $T^3$ dependence of $C_v$ at low temperatures as illustrated by the thermal conductivity of α-quartz along the c-axis (Fig. 2b).

By contrast, the thermal conductivity of a dielectric glass material is much lower and has a markedly different temperature dependence. These features are illustrated by vitreous silica (Fig. 2b) for which $\kappa$ increases non-monotonously with increasing temperature with a remarkable plateau beginning just below the hump in $C_p/T^3$. At very low temperatures, $\kappa$ follows an approximate $T^2$ law instead of the $T^3$ dependence expected from the Debye approximation. According to a widely accepted interpretation, this initial $T^2$ rise is due to interactions between phonons and tunneling states, which reduce $l$ in glasses [7-9].

The microscopic origin of the plateau around 3-15 K is in contrast a much more controversial issue. In early calculations a very efficient phonon scattering mechanism was assumed by a proportionality of $l$ to at least $\Omega^{-4}$ [14], which suggested Rayleigh scattering from disorder in the glassy network. From eqn. (4), it is possible to estimate a mean value for $l(T)$ that can be recast in $l(\Omega)$ with the dominant phonon approximation $\hbar\Omega \sim 3.8\ k_B T$ (Fig. 2b). In the hypersound frequency range, *i.e.* below 100 GHz, $l(\omega)$ exceeds 1 µm so that it is much larger than the acoustic wavelength $\lambda$, confirming the assumption of propagation. In this range, resonant relaxation by tunneling states dominates acoustic attenuation below 1 K (Fig. 2b, top x-scale), yielding $l(\Omega) \propto \Omega^{-1}$. Above about 100 GHz, $l(\Omega)$ drops rapidly following an approximate $\Omega^{-4}$ trend. Near 1 THz, it becomes comparable to the acoustic wavelength $\lambda$, marking the Ioffe-Regel crossover from propagating plane-wave acoustic modes to diffuse excitations. Above the corresponding frequency $\Omega_{IR}$, the wave vector loses its meaning and the notion of phonon becomes ill-defined. Sound waves do not propagate and can no longer transfer energy. The dominant phonon approximation certainly breaks down as well in this range. Only recently has this frequency region become available for coherent spectroscopy of acoustic phonons. At higher temperatures, $\kappa$ rises again and eventually saturates at around 1 W m$^{-1}$ K$^{-1}$. As heat transport can no longer be mediated by propagating sound waves, it is generally admitted that $\kappa$ is governed here by diffusion mechanisms [15,16].

Finally, the room-temperature conductivity of α-cristobalite of ~ 6 W m$^{-1}$ K$^{-1}$ (Fig. 2b) is close to the value obtained in the direction normal to the *c* axis in α-quartz and much higher than the ~ 1 W m$^{-1}$ K$^{-1}$ reported for *v*-SiO$_2$. This feature suggests that the temperature dependence of the thermal conductivity of α-cristobalite is essentially governed by anharmonic processes and follows the expected crystalline increase up to the Casimir limit when the temperature is lowered. Hence, the very rapid flattening of the dispersion curves of the acoustic branches in the <110> direction, which causes the hump in $C_p/T^3$, is by no mean sufficient in itself to produce a plateau in $\kappa$ around 10 K.

## 4. Inelastic spectroscopy in glasses

### 4.1. Dispersion diagram and experimental techniques

The dispersion curves of the longitudinal phonons of vitreous silica and amorphous selenium, in the region where they are plane waves, illustrate the variety of techniques used to study atomic excitations in disordered systems (Fig. 3). The frequency range of network optic modes (typically between few THz and ~2000 THz) is delineated by horizontal dashed lines whereas the boson peak stands on the low-frequency limit of that region (~0.5 - 3 THz). The longitudinal acoustic branch of vitreous silica has for instance been widely investigated by Inelastic X-ray Scattering, whereas *v*-Se is one of the few examples where longitudinal acoustic phonons are sufficiently slow to be accessible by Inelastic Neutron Scattering...

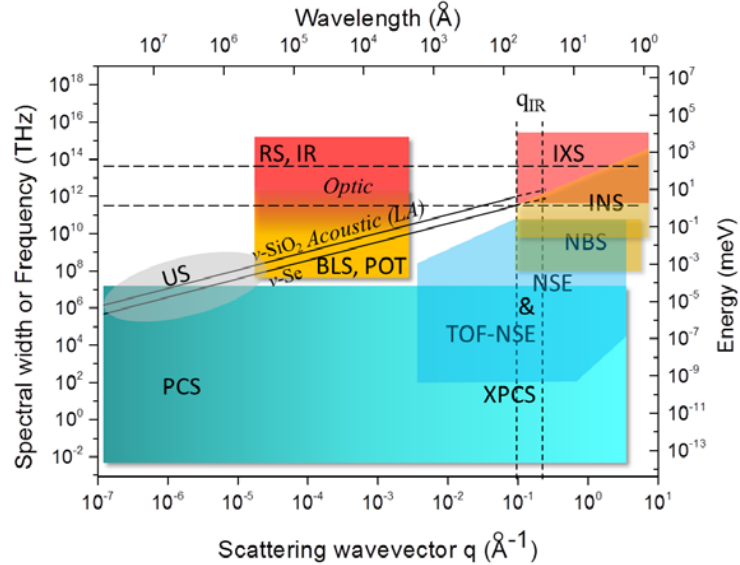

Fig. 3. Dispersion of longitudinal acoustic phonons in vitreous silica ($v_{LA}$ = 5960 m.s−1) and selenium ($v_{LA} \cong 1800$ m.s$^{-1}$) and relevant ranges of the scattering techniques probing atomic excitations in disordered materials: Ultrasonics (US), Brillouin (BLS), Raman (RS), picosecond optical technique (POT), infrared absorption (IR), inelastic neutron (INS) and X-ray scattering (IXS), neutron spin echo (NSE), neutron backscattering (NBS), photon (PCS) and X-ray photon correlation spectroscopy (XPCS). The frequency domain of optic vibrations in network glasses is delineated by horizontal dashed lines.

Depending on their frequency and scattering wavevector, some of techniques displayed in Fig. 3 are sensitive to relaxation only (XPCS, PCS, NSE), others to both relaxation and vibrations (US, BLS, POT, RS, IXS, INS, and NBS) whereas the scattering vector **q** is not defined for IR and US, which do not involve scattering. Much of the dispersion diagram is nowadays experimentally accessible except in the domain between 10 Å$^{-1}$ and ~2 10$^{-3}$ Å$^{-1}$ located at the limit of the Ioffe-Regel crossover region, defined by $q_{IR}$ (section 5.2). Owing to kinematic conditions, Inelastic Neutron Scattering is not a technique of choice for such experiments, at least in structural glasses of rather high sound velocities. Close to q = 0, it is of course not possible to measure sound waves whose velocities are larger than those of the incident neutrons. Accordingly, the high-frequency limit of INS in Fig. 3 roughly corresponds to the highest neutron velocities enabling measurements down to q ~ 0.1 Å$^{-1}$. Another drawback of Inelastic Neutron and X-ray Scattering is their inability to probe transverse acoustic phonons close to q = 0.

*4.2. Scattering intensity*

The intensity **I**(**q**,$\omega$) of an incident radiation scattered by a material is proportional to the space and time Fourier transform of the correlation function of the physical quantity (let's call it *A*), which couples to the incoming radiation [17]:

$$\mathrm{I}(\mathbf{q},\omega) \propto FT\left\{\int_V \int_t A(\mathbf{r},t)A(\mathbf{r}+\mathbf{r}',t+t')\,\mathrm{d}t\mathrm{d}\mathbf{r}\right\} = C(A).S(\mathbf{q},\omega) \qquad (5)$$

For X-ray, neutron, and light scattering, *A* stands for the electronic density $\rho_e$, the coherence length *b* and the dielectric susceptibility $\chi$, respectively, whereas *A* is the derivative of *A* over the atomic displacements *U* of the mode, and describes how the vibration affects the quantity *A*. Accordingly, the function C(*A*) expresses the selection rules of the scattering experiment. It depends on the strength of the coupling of the incident radiation to a vibration and, therefore, modulates the intensity of its spectral shape given by *S*(**q**,$\omega$). For light scattering, the dynamical susceptibility $\chi'$ can be expressed in terms of the polarizability tensor $\bar{\bar{\alpha}}'$, the hyper-polarizability tensor $\bar{\bar{\bar{\beta}}}'$, or the photo-elastic tensor $\bar{\bar{\bar{p}}}'$, giving rise to Raman, hyper-Raman, and Brillouin processes. Neutron scattering is sensitive to all vibrations and enable measuring the phonon dispersion curves over a large (**q**,$\omega$) range. The vibrational density of states is given by the normalized integral over all **q**-values and yields $g(\omega)$. In neutron scattering studies, however, the latter is modulated by the coherent length of the atoms so that it remains an approximate quantity.

However, structural disorder prevents a general theory from being formulated to describe vibrational selection rules hidden in the expression of C(*A*) in glasses, analogous to the role of group theory played for crystals and molecules. This is one of the main reasons why the description of atomic vibrations is hardly accessible. Another limitation is the spatial localization of the modes described below.

*4.3. Coherent and incoherent scattering*

In disordered media a very important consequence of the spatial localization of extended waves is a loss of coherence of the scattering process due to the ill-defined nature of the wavevector **Q** of the vibration. But the momentum keeps conserved when modes are characterized by a spatial extension larger than the probed wavelength, *e.g.* acoustic and optic phonons in crystals

$$\mathbf{Q} = \mathbf{q} = \pm(\mathbf{k}_i - \mathbf{k}_s) \qquad (6)$$

Only the vibrations whose wavevectors **Q** match those of the experiments $\mathbf{q} = \pm(\mathbf{k}_i - \mathbf{k}_s)$ will then be active for a given scattering geometry. Hence, one observes that understanding selection rules requires to distinguish the frequency and wavevector of the vibration ($\Omega$, **Q**) from that accessible by the instrument ($\omega$, **q**).

The other extreme situation corresponds to non-propagating vibrations of fully localized molecular motions as observed in liquids or gases. A perfect localization in real space leads to an infinite spectral broadening of Q, $\Delta Q \rightarrow \infty$, in the Fourier space. Hence, **Q** is not a relevant physical quantity and the vibration scatters at every **q** value.

For the intermediate situation of quasi-localized vibrations, the coherence length of the mode - if it is sufficiently short - may induce a wavevector spectral broadening $\Delta Q$ (horizontal lines in Fig. 1a). In that case, *the contribution at $\omega$ in the vibrational spectrum is the sum over all modes of frequency $\Omega = \omega$ having a wavevector spectral component Q matching the scattering wave-vector q of the experiment.* The above conclusion holds true in glasses for optic modes as well as for short wavelength acoustic waves since the latter progressively transform into quasi-localized vibrating entities at high frequencies (*cf*. section 2).

Another way to address the propagation of acoustic phonons in amorphous solids is to consider plane waves propagating in a mechanically inhomogeneous medium [18] where $\chi'$ depends on position. The disorder induces a spatial modulation of the photoelastic constants, resulting in distorted acoustic waves whose coherence length is limited by local elastic inhomogeneities, as in the model developed in section 2. A full treatment shows that the light-scattering spectrum consists of the usual Brillouin peaks and a background rising as $\omega^2$ originating from the incoherent contribution induced by these local heterogeneities.

## 5. Vibrational spectra

*5.1. Optic modes in the glass formers $SiO_2$ and $B_2O_3$*

Incoherent scattering process combines with structural disorder to produce broad bands in the Raman and infrared spectra. If the glass has a short-range structure similar to that of a crystalline counterpart, its vibrational spectrum will generally represent a smeared out version of that of the crystal. For example, the similarities between vitreous silica and its crystalline counterparts are striking (Fig.4).

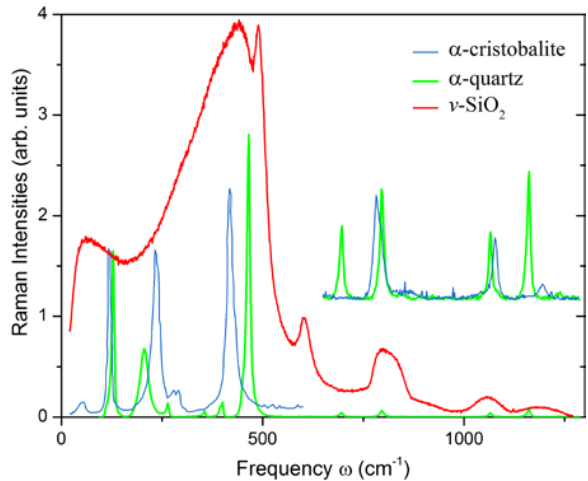

Fig. 4. Raman spectra of $v$-$SiO_2$, $\alpha$-quartz and polycrystalline $\alpha$-cristobalite.

The lack of long-range order prevents vibrations in glasses from being described in a unique way. A first possibility is to consider the atomic-displacements (eigenmodes) of an elementary structural unit such as the $SiO_4$ tetrahedron of $T_d$-symmetry or the Si-O-Si bridge of $C_{2v}$-symmetry in $v$-$SiO_2$, and the $BO_3$ triangle of $D_{3h}$-symmetry or the B-O-B bridge in $v$- $B_2O_3$. A second possibility is to define rocking, bending and stretching motions of N-O-N units, where N=Si, B, Ge, etc. A pure bending modulates only the N-O-N bond angle whereas a pure stretching modifies the N-O bond length. Finally, numerical simulations often project displacements over three orthogonal axes also defined as bending, stretching and rocking axes. The first one is perpendicular to the N-O-N plane, the second is parallel to the N-O-N bisector, and the third is parallel to the N-N direction. In that case, bending and stretching are not pure vibrational modes. These definitions are not unequivocal, however, so that they can foster confusion in mode assignments. Owing to network connectivity, the spectral responses generally involve several structural units, which further complicates the description. Although some of the conclusions are still discussed, atomic displacements are tentatively summarized below for the main glass formers.

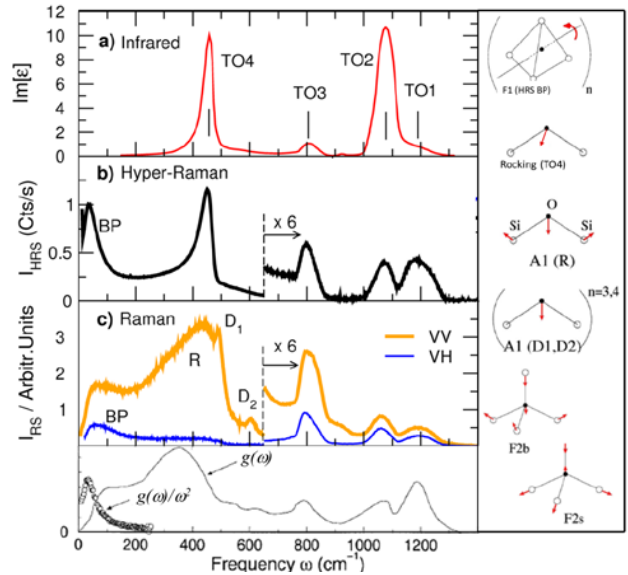

Fig. 5. Vibrational spectroscopy of $v$-$SiO_2$ and atomic displacements for the main modes [19-21]. From top to bottom: Infrared absorption, hyper-Raman scattering, Raman scattering, and vibrational density of states derived from neutron scattering [22]. N.B. $F_1$ vibrations seen in HRS are not the only ones participating in the boson peak.

For $v$-$SiO_2$ and $v$-$B_2O_3$, the differences between the IR, Raman and hyper-Raman spectra (Figs. 5 and 6) indicate that selection rules do apply in glasses with sufficiently well-defined local structures. The three polar modes TO1, TO2, TO3 in $v$-SiO$_2$ are for example active in IR, Raman and hyper-Raman, in agreement with either the $C_{2v}$- or $T_{\underline{d}}$-symmetry selection rules. Within the T$_d$ point group, $F_2$-symmetry stretching motions of SiO$_4$ tetrahedra ($F_{2s}$) contribute preferentially to TO1 and TO2, whereas $F_2$-symmetry bending ones ($F_{2b}$) dominate in TO3 [20]. Since the Raman inactivity of TO4 cannot be accounted for by any of the internal vibrations proposed by the two structural models, this mode rather represents highly cooperative motions involving rocking of the oxygen atoms [21,23] in the Si-O-Si bridges. The weak depolarization ratios ($I_{VH}/I_{VV}$) of the Raman $R$, $D_1$, and $D_2$ bands are compatible with $A_1$-bending of the Si-O-Si bridges of $C_{2v}$-symmetry. The displacements are in-phase in the three- and four-membered rings (Si-O-Si)$_n$ planar structures ($n=3$ and 4, respectively), and are commonly ascribed to breathing modes. Such motions induce very weak dipolar fluctuations and are, therefore, almost inactive in IR and HRS. Likewise, in the vibrational spectrum of $v$-B$_2$O$_3$ (Fig. 6) the atomic displacements proposed are compatible with the molecular selection rules of the $D_{3h}$-symmetry group of BO$_3$ triangles and B$_3$O$_6$ boroxol rings.

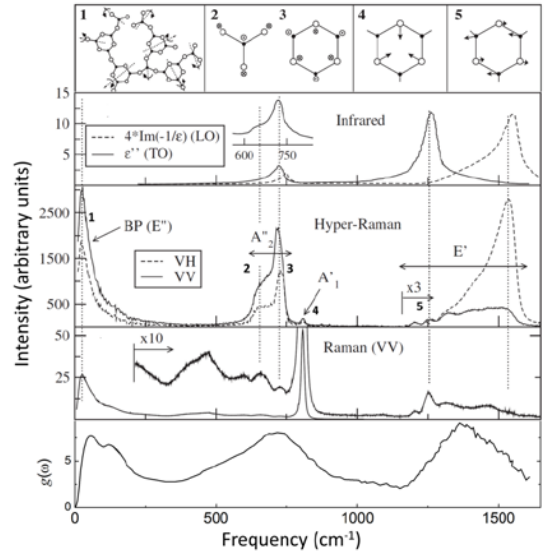

Fig. 6. Vibrational spectroscopy of $v$-SiO$_2$ and atomic displacements for the main modes [19-21]. From top to bottom: Infrared absorption, hyper-Raman scattering, Raman scattering, and vibrational density of states derived from neutron scattering [22]. N.B. $F_1$ vibrations seen in HRS are not the only ones participating in the boson peak.

Although very powerful, vibrational techniques do not lend themselves to quantitative structural estimates. Even though numerical simulations are proving valuable in this respect thanks to calculations of the coupling-to-light coefficients $C(A)$ in eqn. (5) or to RMN calibrations, the main asset of inelastic spectroscopies remains their ability to derive information that may be difficult or even impossible to obtain with conventional techniques. For example, three- and four-membered rings represent only 1 per 670 and 550 $SiO_2$ units, respectively [24]. These proportions are much too low to be detected with diffraction techniques, but high enough to produce the narrow $D_1$ and $D_2$ Raman bands. Whereas the broad $R$-band probes the Si-O-Si angle distribution [22,25], the high-frequency feature depends on the fraction of $Q^n$-species in binary or more complex silicates. The role of modifier, charge compensator or glass former played by other cations can also be investigated. In $v$-$B_2O_3$, structural information accessible by vibrational techniques concerns for instance the fraction of boroxol rings $B_3O_6$ and the ratio of $BO_3$-triangles and $BO_4$-tetrahedra in binary glasses (Ch. 2.8, 7.6).

*5.2. Acoustic excitations*

At large wavelength, *i.e.*, in the continuum limit, acoustic waves propagate in glasses as they do in crystals. Almost non-dispersive sound waves have been indeed measured at low frequencies with ultrasonic (MHz) Brillouin scattering (GHz) or picosecond optical pump-and-probe (100 GHz) experiments. At these frequencies, quasi-plane waves do propagate but with higher attenuation rates than generally found in crystalline solids. The anharmonicity of thermal atomic vibrations causes the attenuation of sound waves in crystals [26] and is in addition reflected in the temperature dependence of the sound velocities or in thermal expansion. The same mechanism is also present in glasses, but it coexists with other damping processes that are mostly specific to them.

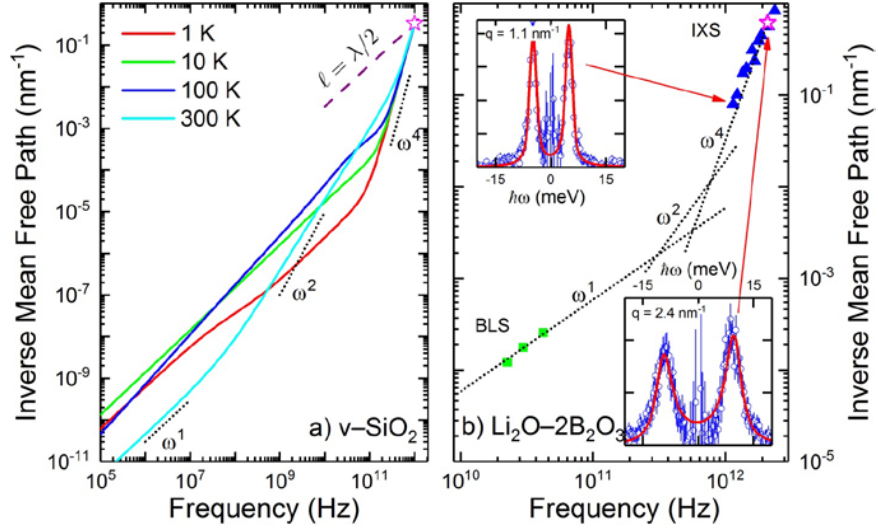

Fig. 7. Inverse mean free path $l^{-1}(\omega,T)$ of longitudinal acoustic excitations in glasses. (a) For $v$-SiO$_2$: curves calculated with four damping mechanisms, namely, interactions with tunneling states, thermally activated relaxation, anharmonicity and Rayleigh scattering at high frequency, which leads to the Ioffe-Regel crossover at ~1 THz (star). (b) For lithium diborate glass as measured at 573 K with BLS (green squares) and IXS (filled triangles). Dotted lines referred to the different damping processes: thermally activated relaxation plus lithium diffusion at GHz frequencies, anharmonicity around 0.5 THz and Rayleigh scattering at THz frequencies. IR-crossover at ~2.4 THz (star). Inset: dramatic increase of the Brillouin linewidth in IXS spectra in the IR crossover between q =1.1 and 2.4 nm$^{-1}$.

At low frequencies, the energy mean free path of acoustic phonons is limited by resonant interactions with tunneling states at low temperatures and by thermally activated relaxation at higher temperatures, both leading to $l^{-1} \propto \omega$ (Fig. 7a). The latter mechanism is empirically described by the relaxation of group of atoms or "defects" between two or more equilibrium positions [27,28]. At hypersonic frequencies, anharmonic processes coexist with thermal activated relaxation to drive a smooth transition from a $l^{-1} \propto \omega$ law to a $l^{-1} \propto \omega^2$ trend from low to high temperatures as clearly observed above GHz frequencies (Fig. 7a). The combination of these mechanisms thus results in complex variations of $l^{-1}(\omega,T)$ which strongly depend on glass composition. Finally, very short-wavelength sound waves experience a disorder-induced strong scattering regime, which dominates above 0.1-0.2 THz, depending on temperature and on the efficiency of the other damping mechanisms. The dramatic increase of $l^{-1}(\omega)$ rapidly produces the Ioffe-Regel crossover at the frequency $\omega_{IR}/2\pi$ (Fig. 7a). Such a rapid decrease of the mean free path is actually the trend required to produce the aforementioned plateau in $\kappa(T)$ around 10 K.

With techniques ranging from Brillouin to inelastic X-ray scattering, it has become possible to study longitudinal acoustic excitations just below the IR-crossover. In two favorable cases [29,30], detailed measurements have revealed the existence of a very rapid increase of the Brillouin linewidth $\Gamma = l^{-1}v_L$, where $v_L$ is the longitudinal sound velocity as illustrated by lithium diborate glass [$Li_2O \cdot 2B_2O_3$] at 573 K (Fig. 7b, inset). In the Brillouin spectrum taken at the smallest usable Q, the fitted elastic peaks have been subtracted from both the data and the fit to make the inelastic parts and their damped harmonic oscillator components apparent. At this low-q value, the Brillouin width is almost entirely instrumental, whereas it contains a real and large broadening in the bottom inset. At that point, the lineshape of the damped harmonic oscillator begins to deviate increasingly from the measured signal. The reciprocal of the energy mean free path corresponding to the Brillouin linewidths from all the measured IXS spectra are reported in the main graph of Fig. 7b.

Following a $\omega^4$ trend, the dramatic increase of $l^{-1}$ is clearly evidenced over a decade. At q = $q_{IR}$ = 2.4 nm$^{-1}$, the Brillouin linewidth reaches one-third of the Brillouin frequency, which is the Ioffe-Regel criterion. The IXS spectra measured at higher q values are inconsistent with a single damped-harmonic oscillator response, demonstrating that this approximation is no longer valid, thus marking the end of acoustic plane waves at $q_{IR}$ in this glass. In Brillouin light-scattering experiments made at the same temperature (Fig. 7b), the $l^{-1}$ values obtained also show that damping in this ($\omega,T$) region is dominated by thermally activated relaxation processes and lithium diffusion, leading to an approximate $l^{-1} \propto \omega$ dependence, in agreement with ultrasonic experiments. From the temperature variation of the sound velocity, an estimate of the anharmonicity contribution at 573 K ($l^{-1}_{anh} \propto \omega^2$ in Fig. 7b) shows that this mechanism should dominate in a narrow frequency range around the crossover between $\omega^1$ and $\omega^4$ trends. More recently, similar IXS measurements made on $v$-SiO$_2$ and glycerol [$C_3H_8O_3$] have also indicated such a dramatic increase of the reciprocal of the mean free path of longitudinal acoustic excitations at THz frequencies.

The results obtained for a densified *v*-SiO$_2$ and then on lithium diborate glass have represented the first experimental evidence for the existence of a Rayleigh-type scattering mechanism of acoustic phonons in glasses producing an Ioffe-Regel crossover at some large acoustic wavelengths. The observation of both phenomena firmly demonstrated that the plane-wave approximation has already lost all validity at an intermediate scale, *i.e.*, at about 1-3 nm depending on the glass. The observed crossover from propagating to diffusing sound waves has also offered a natural explanation for the anomalous low-temperature plateau in thermal conductivity. In several glasses investigated so far, $\omega_{IR}$ is close to the boson peak frequency $\omega_{BP}$, suggesting a direct relationship between both quantities. The correspondence is not one to one, however, as $\omega_{IR}$ remains at or above $\omega_{BP}$. It is believed that the boson peak frequency actually corresponds to the Ioffe-Regel crossover frequency for the transverse acoustic excitations in all glasses, but no experimental data have demonstrated the validity of this statement yet.

## 6. The boson peak

The excess of vibrational modes in the reduced vibrational density of states $g(\omega)/\omega^2$ translates into a broad and asymmetric band at frequencies between 0.5 and 3 THz in Raman or infrared spectra, corresponding approximately to the end of the acoustic branches. Vibrations of different origins are likely participating in that frequency region and should be separated into glass-specific vibrations (*boson peak modes*) and crystalline-like vibrations (non-boson peak modes). Many attempts have been made to relate the boson peak *e.g.* to the medium range order [31], the crystalline phases [11], or the macroscopic properties [32]. This section will focus on the origin of the vibrations underlying this excess of modes.

*6.1 Oxide glasses*

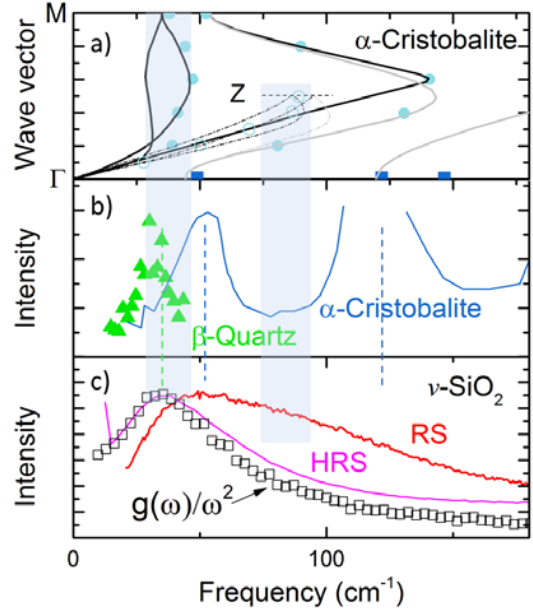

Fig. 8. Vibrational excitations in *v*-SiO$_2$. (a) Dispersion curve of acoustic modes and low-lying optic branches along the $\Gamma \rightarrow M$ (plain lines) and $\Gamma \rightarrow Z$ (dot dashed lines) directions of the Brillouin zone; squares Raman data. (b) Raman spectra of α-cristobalite and neutron spectra of the soft mode of β-quartz at 1250 K whose position at $T_g$ is indicated by the dashed line at around 36 cm$^{-1}$. (c) Spectroscopic signatures of the boson peak; possible vibrational contributions indicated by shadowed regions and the vertical dashed lines.

In quartz, a very important vibration is the soft mode associated with the structural α-β instability at 846 K whose displacements have been interpreted as librations of SiO$_4$ tetrahedra. Its temperature dependence has been measured by neutron scattering (Fig. 8b) and its frequency extrapolates to ~36 cm$^{-1}$ at $T_g$~1600 K, *i.e.*, to the frequency of the maximum of $g(\omega)/\omega^2$ measured by inelastic neutron scattering (Fig. 8c). Librations of rigid units are also soft vibrations in β-cristobalite but are located at the zone boundary of a transverse acoustic branch. These low-frequency excitations have been identified as *rigid unit modes* [33], *i.e.*, external modes combining librations and translations of rigid elementary units. These units associate with weak inter-unit restoring forces in the glass and therefore vibrate at low frequency.

Similarly, inelastic neutron scattering, hyper-Raman, as well as numerical simulations have highlighted the presence of librations of rigid $SiO_4$ ($F_1$-symmetry displacements in Fig. 5) at boson-peak frequencies in vitreous silica [34,35]. Librations also participate at boson-peak frequencies in pure boron oxide (Fig. 6) but with a lower weight than in $v$-$SiO_2$. In this respect, silica is probably a peculiar system since librations are soft modes of the crystalline polymorphs, which to our knowledge is not the case for boron oxide. In connected networks, librations couple with translations of rigid units and hybridize with transverse and longitudinal acoustic phonons of similar frequency. This source of scattering combines with the strong scattering process of sound waves (shadowed regions in Figs. 8a-c) to built up the reservoir of *boson peak modes*. The loss of the Brillouin zone combined with the localization of the plane waves in the glass makes identification of these vibrations as acoustic- or optic-like useless. Owing to the different nature of these modes, the boson peak manifests itself in a way specific to the spectroscopic technique used, leading to differences in neutron, Raman, hyper-Raman, and infrared responses. What matters first and foremost is that these excitations relate to the structural disorder of the glass and to the plateau of the thermal conductivity.

*Crystalline-like modes* may also contribute at frequencies beyond that of the boson peak such as the Raman modes at ~50 cm$^{-1}$ and ~120 cm$^{-1}$ in α-cristobalite (Fig. 8b) [36] or cation modes in soda-lime-silicate glasses [37]. Indeed, the vibrational density of states $g(\omega)$ of $v$-$SiO_2$ below 300 cm$^{-1}$ (as well as its Raman spectra) is very complex, pointing to numerous low-frequency optic vibrations (Fig. 5). Conversely, that of $B_2O_3$ looks like Van-Hove singularities of transverse and longitudinal acoustic branches (Fig. 6) emphasizing thereby a much weaker contribution from *crystalline-like* optic modes.

## 6.2. Other glasses

The picture is even more complicated in molecular glasses (*e.g.* glycerol, ortho-terphenyl, etc.) where a large number of intramolecular and reorientational motions are likely to contribute at low-frequency, leading to a great complexity in both vibrational responses and structural relaxation (Ch. 8.6). Hence, these systems generally do not represent good model systems for studying the boson peak.

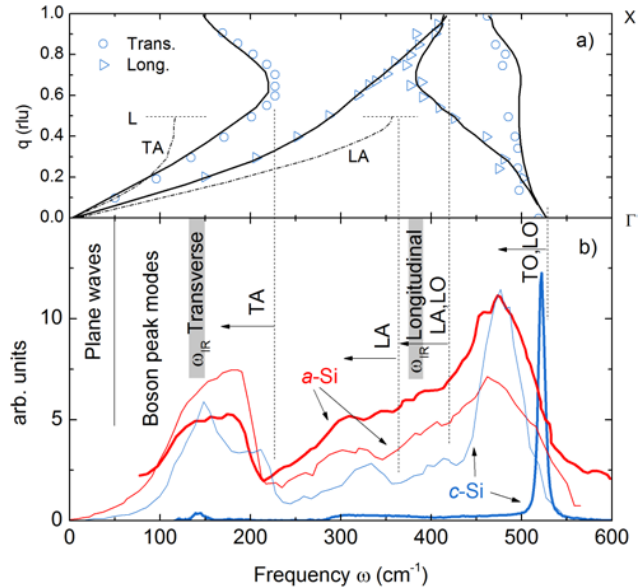

Fig. 9. Vibrational excitations in crystalline and amorphous silicon. (a) Dispersion curves along the X (symbols and lines) and L (dot-dashed lines) directions of the crystal [38]; dispersion curves of acoustic phonons about 25 % lower in the glass than in the crystal. (b) Vibrational density of states $g(\omega)$ (thin lines) [41] and Raman scattering spectra (bold lines) of amorphous silicon [37] and $c$-Si. Plane wave, boson peak mode mode, and $\omega_{IR}$ transverse and longitudinal acoustic limits taken from [16].

Metallic glasses may in contrast prove useful in this respect. The simplest ones can be usually described as chemically and positionally disordered dense-packed structures without orientational interactions or intramolecular processes (Ch. 7.10). Consistent with the Lennard-Jones interacting-sphere model, the excess of low frequency modes in $C_p$ between ~10 and 30 K is then often composed of purely harmonic vibrations, *i.e.,* of Einstein modes corresponding to local excitations of loose atoms in the glassy structure. Unfortunately, electronic and atomic properties often mix in metallic glasses, which may confuse the description of the boson peak and thermal properties. Electronic processes at impurity sites may for example contribute to localized excitations at boson-peak frequencies whereas

24thermal conductivity may also be dominated by electronic heat-transfer channels, which are much more efficient than atomic ones in conductors.

In the large family of non-insulating glasses, amorphous silicon (*a*-Si) is probably one of the simplest examples. Its crystalline counterpart (*c*-Si) has only one triply-degenerate optic mode at $\omega = 520$ cm$^{-1}$, which leads to an interesting situation where only acoustic branches contribute to the vibrational density of states below ~350 cm$^{-1}$ (Fig. 9a). The crystalline tetrahedral short-range order is preserved in the glassy state so that the densities of the two forms differ by 1.8 % only. This structural similarity lends support to qualitative comparisons between vibrational properties. For example the vibrational densities of states are similar below ~350 cm$^{-1}$ with only a down-shift of the band maxima in the glass, reflecting lower sound velocities (Fig. 9b). Despite the absence of optic modes at low frequency, the Raman spectrum is not flat but displays broad structures that have been associated with both second-order processes on acoustic branches [38,39] and disordered-induced scattering due to acoustic plane-wave destruction (section 2 and 5.2). The latter effect seems to be rather weak in *a*-Si, however, since it yields a moderate damping mechanism of the acoustic phonon branches, proportional to $q^2$ [16,40], as compared with the very fast $q^4$-law observed in network glasses. This probably explains why the Raman response is very similar to the $g(\omega)$ measured by Inelastic Neutron Scattering (Fig 9b) [41].

As in crystalline silicon, $C_p/T$ in *a*-Si varies without showing evidence for two-level systems at least down to 2 K [4,42]. The weak damping regime of the acoustic phonons goes against the formation of a plateau in the thermal conductivity as well. Unfortunately, however, existing reports are controversial and firm experimental evidence is still lacking on that question. Finally, the Ioffe-Regel crossover frequency is upshifted toward the end of the transverse acoustic branch (Fig 9b) [16], which raises some doubts about its physical meaning. In the current state of knowledge, the only reliable glass-specific vibrational feature of amorphous silicon is, therefore, the partial destruction of acoustic plane-waves at large *q* (*i.e.*, short wavelengths). Although moderate, this mechanism generates *boson-peak modes* at frequencies between the plane-wave regime and the Ioffe-Regel crossover at $\omega_{IR}$.

**7. Perspectives**

Inelastic scattering selection rules in glasses are relaxed by local structural and mechanical disorder, yielding broad and asymmetric spectral responses. Unlike for crystals, there exists no well-established analytical theory of the atomic displacements that underlie the spectral responses. Analyses remain mostly phenomenological, except perhaps for simple glasses for which atomistic simulations nowadays provide a firmer theoretical foundation. Indeed the nature of the vibrations in the main oxide glass formers, up to binary and to a lesser extent ternary systems, are relatively well understood. Extracting quantitative structural information, in particular from RS, remains a challenging issue.

Even more challenging is the nature of the boson peak. Its modes participate at low frequency in the vibrational spectra of most disordered systems. These vibrating entities mostly originate from a renormalization and a redistribution of the modes of the acoustic branches due to the destruction of plane waves of nanometer wavelengths. They can eventually hybridize with low-lying optic vibrations (rigid-unit modes), such as the librations of rigid $SiO_4$ tetrahedra in silica or loose local atomic motion in Lennard-Jones type metallic glasses. The purely acoustic-type damping mechanism, enhanced by the second one when relevant, induces a very fast decay of the mean free path of the acoustic phonons at THz frequency, as evidenced by IXS and numerical simulations. There follows a crossover from a propagative to a diffusive character of sound waves and offers a natural explanation of the low-temperature plateau of thermal conductivity.

These modes build up the boson-peak structure mostly below its maximum in $g(\omega)/\omega^2$ plots in Raman and inelastic neutron scattering spectra. At higher frequencies, this spectral response mixes with those of other mechanisms, such as incoherent scattering from high-frequency acoustic branches due to the loss of wavevector selection rules, direct scattering of low-lying

26optic branches, second- or higher-order Raman scattering processes, and possibly others. It is tempting to define the boson peak as arising from the sum of all the excitations that construct the broad feature at THz frequency in glass, as it is often done in literature.

Here we have preferred to separate the *boson peak modes* from the other scattering processes since the latter are related to spectroscopic considerations whereas the former accounts for solid-state properties specific to glasses. Current advances in numerical simulations have allowed the boson peak modes at the origin of the plateau in $\kappa$ to be identified in very few cases [16,43]. Highlighting them over the other scattering channels in a scattering experiment probably constitutes one of the most challenging issues of the coming decades.

**Acknowledgments**

P. Gujrati and R. Vacher are gratefully tanked for their helpful comments on this chapter.

**Appendix** : **supplementary information for figure captions**

**Fig. 2.** Data from

F. Birch and H. Clark, *Amer. Jour. Sci.,* 238 (1940) 529-58.

U. Buchenau, M. Prager, N. Nücker, A.J. Dianoux, N. Ahmad and W.A. Phillips, *Phys. Rev. B*, 34 (1986) 5665-73.

D.G. Cahill and R.O. Pohl, *Phys. Rev. B*, 35 (1987) 4067-73.

A. Eucken. *Annalen der Physik*, 34 (1911) 185-221.

P. Flubacher, A.J. Leadbetter, J.A. Morrison and B.P. Stoicheff, *J. Phys. Chem. Solids*, 12 (1959) 53-58.

**Fig. 5.** Data from

**Fig. 8.** Data from